\documentclass[preprint,journal]{vgtc}   
\usepackage{amsmath,amsfonts}
\usepackage{algorithmic}
\usepackage{algorithm}
\usepackage{array}
\usepackage{textcomp}
\usepackage{stfloats}
\usepackage{url}
\usepackage{verbatim}
\usepackage{graphicx}
\usepackage{cite}
\usepackage{comment}
\usepackage{soul}
\usepackage{transparent}
\usepackage[svgnames]{xcolor}
\usepackage{tcolorbox}
\usepackage{setspace}
\usepackage{tikz}
\usepackage{fancyhdr}
\usetikzlibrary{calc}
\usepackage{wrapfig}
\usepackage{endnotes}
\definecolor{OliveGreen}{RGB}{183, 207, 178}
\definecolor{DarkYellow}{HTML}{DEA601}
\definecolor{DarkOrange}{HTML}{ED7D31}
\definecolor{DarkBlue}{HTML}{4472C4}
\usepackage{xspace}
\newcommand{\taskWhy}{\textit{\textbf{\textcolor{DarkOrange}{Why?}}}\xspace}
\newcommand{\taskHow}{\textit{\textbf{\textcolor{DarkBlue}{How?}}}\xspace}
\newcommand{\taskWhat}{\textit{\textbf{\textcolor{DarkYellow}{What?}}}\xspace}

\usepackage[compact]{titlesec}
\titlespacing{\section}{0pt}{1ex}{1ex}
\titlespacing{\subsection}{0pt}{1ex}{1ex}

\newcommand{\drawCircle}[3]{\tikz\draw[#1,fill=#2] (0,0) circle (#3);}

\renewcommand{\paragraph}[1]{\textbf{#1:}}

\definecolor{indvHigh}{HTML}{9BB2E0}
\definecolor{indvMed}{HTML}{C8D6ED}
\definecolor{indvLow}{HTML}{ECF2F9}

\definecolor{grpHigh}{HTML}{B6E09B}
\definecolor{grpMed}{HTML}{D6EDC7}
\definecolor{grpLow}{HTML}{F2F9ED}

\definecolor{ensTaskHigh}{HTML}{B86029}
\definecolor{ensTaskMed}{HTML}{EAB38B}
\definecolor{ensTaskMedLow}{HTML}{EFCAB0}
\definecolor{ensTaskLow}{HTML}{F8E4D8}

\definecolor{ensTotHigh}{HTML}{B7912F}
\definecolor{ensTotMed}{HTML}{F5C242}
\definecolor{ensTotMedLow}{HTML}{F8DA78}
\definecolor{ensTotLow}{HTML}{FDF3D0}

\definecolor{dsHigh}{HTML}{2F5597}
\definecolor{dsMed}{HTML}{8FAADC}
\definecolor{dsLow}{HTML}{B4C7E7}

\onlineid{1295}

\preprinttext{To appear in IEEE Transactions on Visualization and Computer Graphics}

\vgtccategory{Research}

\vgtcpapertype{theory/model}

\title{A Qualitative Analysis of Common Practices in Annotations: A Taxonomy and
Design Space}

\author{%
  \authororcid{Md Dilshadur Rahman}{0009-0008-5467-615X},
  \authororcid{Ghulam Jilani Quadri}{0000-0002-8054-5048}, 
  Bhavana Doppalapudi,
  Danielle Albers Szafir,
  \authororcid{Paul Rosen}{0000-0002-0873-9518}
}

\authorfooter{
  \item Md Dilshadur Rahman\textsuperscript{1} and Paul Rosen\textsuperscript{2} are with the University of Utah. E-mails: dilshadur@sci.utah.edu\textsuperscript{1} and paul.rosen@utah.edu\textsuperscript{2}
  \item Ghulam Jilani Quadri is with the University of Oklahoma. E-mail: quadri@ou.edu
  \item Bhavana Dopplapudi is with the University of South Florida. E-mail: bdoppalapudi@usf.edu
  \item Danielle Albers Szafir is with the University of North Carolina at Chapel Hill. E-mail: danielle.szafir@cs.unc.edu
}

\abstract{%
Annotations play a vital role in highlighting critical aspects of visualizations, aiding in data externalization and exploration, collaborative sensemaking, and visual storytelling. However, despite their widespread use, we identified a lack of a design space for common practices for annotations. In this paper, we evaluated over 1,800 static annotated charts to understand how people annotate visualizations in practice. Through qualitative coding of these diverse real-world annotated charts, we explored three primary aspects of annotation usage patterns: analytic purposes for chart annotations (e.g., present, identify, summarize, or compare data features), mechanisms for chart annotations (e.g., types and combinations of annotations used, frequency of different annotation types across chart types, etc.), and the data source used to generate the annotations. We then synthesized our findings into a design space of annotations, highlighting key design choices for chart annotations. We presented three case studies illustrating our design space as a practical framework for chart annotations to enhance the communication of visualization insights. All supplemental materials are available at \url{https://shorturl.at/bAGM1}.

}


\teaser{
    \centering
        \begin{minipage}[b]{0.98\linewidth}
        \includegraphics[trim=2pt 2pt 5pt 2pt, clip, width=\linewidth]{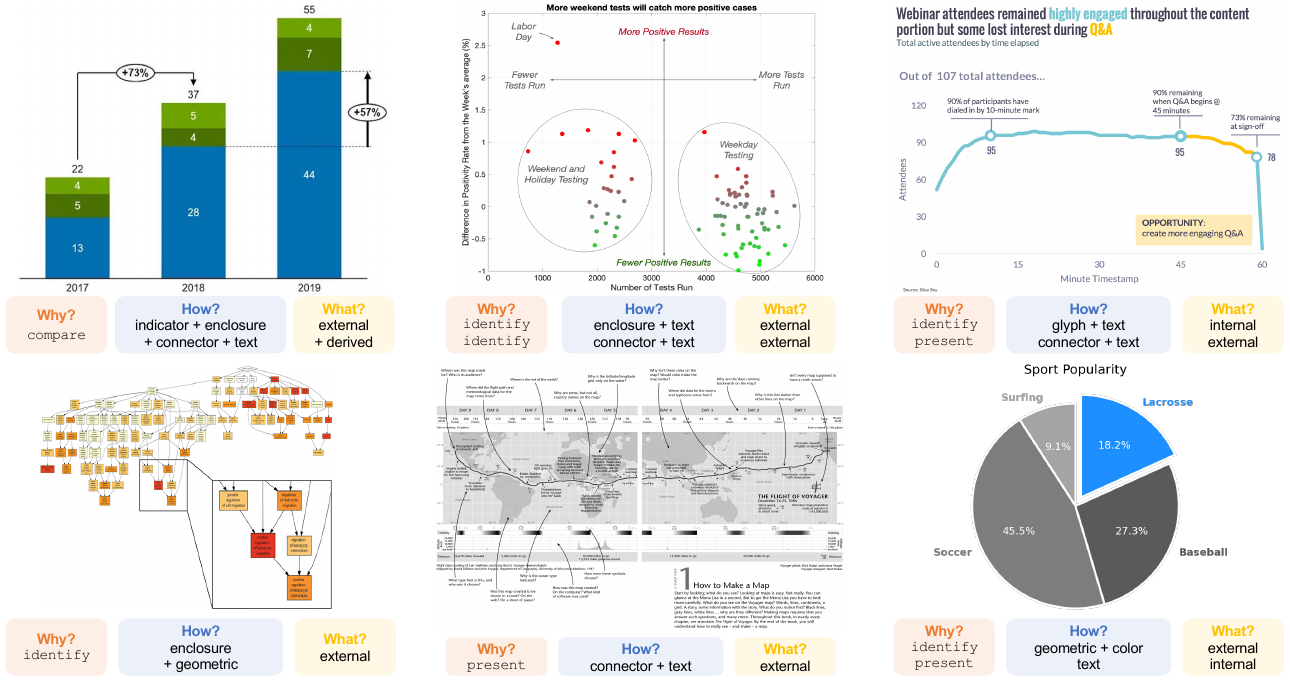}
    \end{minipage}
    \begin{minipage}[b]{1pt}
        \hspace{-465pt}
        \begin{minipage}[t][8pt][c]{6pt}\drawCircle{black}{LightGray}{5pt}\end{minipage}
        \hspace{140pt}
        \begin{minipage}[t][8pt][c]{6pt}\drawCircle{black}{LightGray}{5pt}\end{minipage}
        \hspace{144pt}
        \begin{minipage}[t][8pt][c]{6pt}\drawCircle{black}{LightGray}{5pt}\end{minipage}
        
        \vspace{-14pt}
        \hspace{-462.5pt}
        \subfloat[\label{fig:teaser:a}]{}
        \hspace{148.8pt}
        \subfloat[\label{fig:teaser:b__old_d}]{}
        \hspace{153.0pt}
        \subfloat[\label{fig:teaser:c}]{}
        
        \vspace{110pt}
        \hspace{-465pt}
        \begin{minipage}[t][8pt][c]{6pt}\drawCircle{black}{LightGray}{5pt}\end{minipage}
        \hspace{130pt}
        \begin{minipage}[t][8pt][c]{6pt}\drawCircle{black}{LightGray}{5pt}\end{minipage}
        \hspace{180pt}
        \begin{minipage}[t][8pt][c]{6pt}\drawCircle{black}{LightGray}{5pt}\end{minipage}
        
        \vspace{-14pt}
        \hspace{-462.5pt}
        \subfloat[\label{fig:teaser:d_old_b}]{}
        \hspace{138.8pt}
        \subfloat[\label{fig:teaser:e}]{}
        \hspace{189.3pt}
        \subfloat[\label{fig:teaser:f}]{}
        \vspace{85pt}
    \end{minipage}

    \caption{Examples of annotated charts collected from the internet and analyzed within our annotation design space, which is organized around \taskWhy (annotation tasks), \taskHow (types of annotations), and \taskWhat (source of annotation data). (a)~The bar chart uses \textit{connector}, \textit{indicator}, \textit{enclosure}, and \textit{text} annotations to help \texttt{compare} between years; (b)~the scatterplot uses \textit{text}, \textit{connector}, and \textit{enclosure} annotations to assist in \texttt{identifying} individual or groups of data points; (c)~the line chart uses \textit{glyph} and \textit{text} annotations to aid \texttt{identify} points of interest, and \textit{text} and \textit{connector} annotations to \texttt{present} key events; (d)~the node-link diagram utilizes an \textit{enclosure} and a \textit{geometric} annotation to help \texttt{identify} the area of interest; (e)~the map has \textit{connector} and \textit{text} annotations to \texttt{present} key events on the journey and critiquing the visualization; and (f)~the pie chart has \textit{text} to \texttt{present} exact values from the dataset and \textit{color} and \textit{geometric} annotations to support \texttt{identifying} the lacrosse slice.}
    \label{fig:teaser}

}

\graphicspath{{figs/}{figures/}{pictures/}{images/}{./}} %

\usepackage{mathptmx}                  %

\begin{document}

\maketitle

\begin{figure*}[!t]
    \centering

    \begin{minipage}[b]{0.95\linewidth}
        \includegraphics[width=\linewidth]{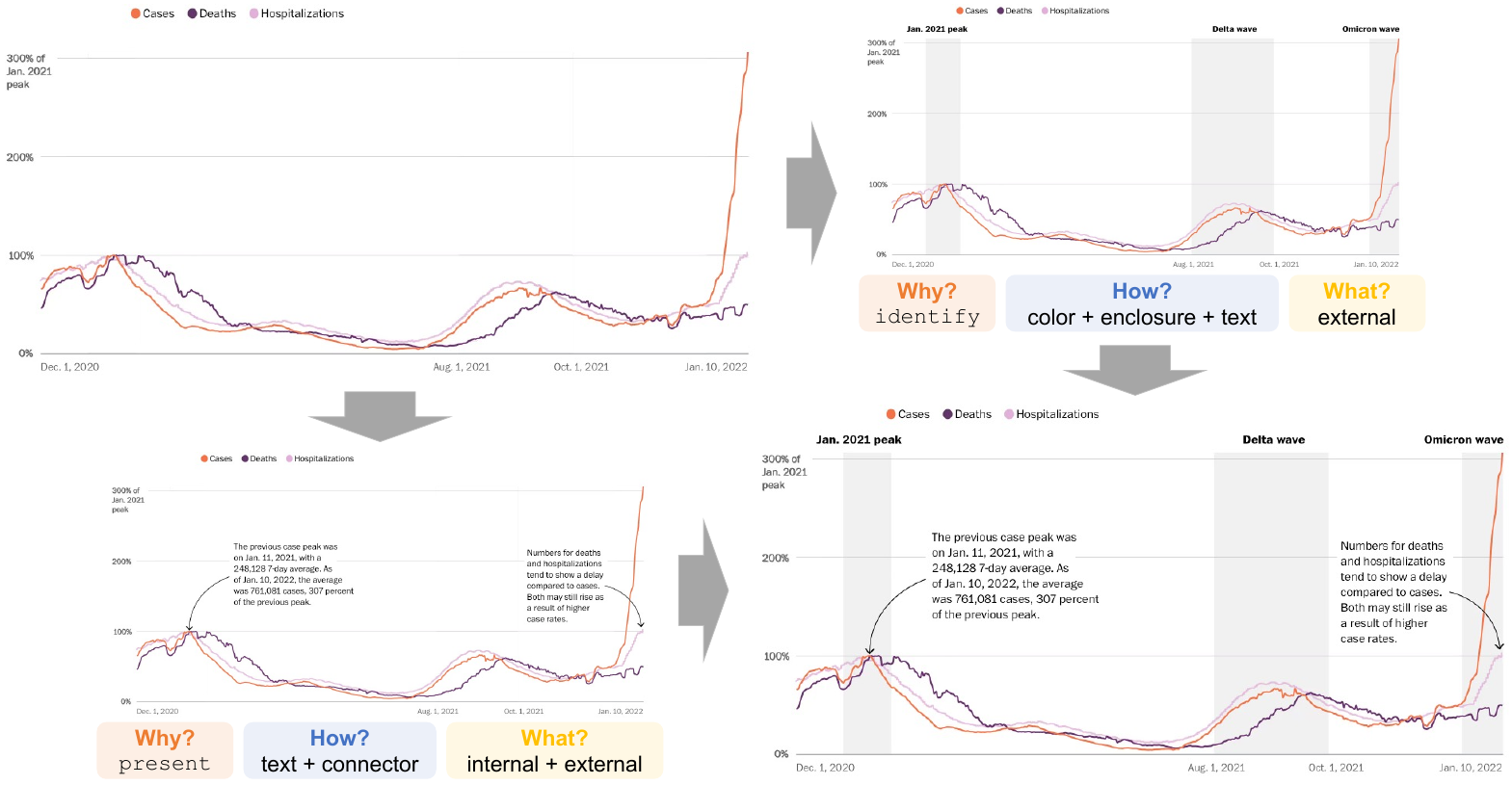}
    \end{minipage}
    \begin{minipage}[b]{1pt}
        \hspace{-498pt}
        \subfloat[\label{fig:cs3-a}]{\hspace{20pt}}
        \hspace{440pt}
        \subfloat[\label{fig:cs3-b}]{\hspace{20pt}}
        \vspace{105pt}
    
        \hspace{-468pt}
        \subfloat[\label{fig:cs3-c}]{\hspace{20pt}}
        \hspace{440pt}
        \subfloat[\label{fig:cs3-d}]{\hspace{20pt}}
        \vspace{95pt}

    \end{minipage}

    \vspace{-5pt}
    \caption{A line chart from \textit{The Washington Post}~\cite{Tan2022} illustrates COVID-19 peak comparisons, plotting time on the horizontal axis and percentage growth relative to the January 2021 peak vertically: (a)~shows the baseline chart with basic visualization elements (i.e., axes, labels, lines, legends, and gridlines) but with annotations removed; (b)~uses color+enclosure+text ensembles of annotations to help \texttt{identify} the peaks of different COVID-19 waves; (c)~uses text+connector ensembles to \texttt{present} additional context from the associated article; and (d)~displays the completely annotated chart.}
    \label{fig:case-study-3}
\end{figure*}

\section{Introduction}
\label{sec.introduction}
Annotations, as supplementary graphical or textual elements~\cite{munzner2014visualization}, highlight key features in visualizations and provide additional context to the data. They facilitate the mental organization of information for viewers, enhancing the memorability and recall of data presented in visualizations~\cite{cedilnik2000procedural, borkin2015beyond, borkin2013makes, bateman2010useful, ajani2021declutter, kong2019understanding}. Annotating is also considered one of the critical tasks in interactive visual analysis~\cite{heer2012interactive, zhao2016annotation, taher2015exploring}. Moreover, annotations play various critical roles in visualizations, aiding in externalizing and exploring data~\cite{mahyar2012note, shrinivasan2009connecting, kim2019inking}, supporting collaborative sensemaking~\cite{mahyar2014supporting,isenberg2006interactive, ellis2004collaborative}, and enhancing narrative storytelling~\cite{segel2010narrative, lee2015more, hullman2011visualization}.

Considering the significance of annotations in visualizations, understanding their design space becomes crucial for creating visualizations. Prior works in this area have focused on limited chart and annotation types or a specific domain of annotation usages, overlooking a thorough analysis of annotations in broad contexts~\cite{ren2017chartaccent,hullman2013contextifier}. For example, Ren et al.'s study in \textit{ChartAccent} explored textual and graphical annotations in limited chart types (i.e., bar charts, line charts, and scatterplots) for visual storytelling~\cite{ren2017chartaccent}, whereas Hullman et al.'s \textit{Contextifier} research focused on narrative visualizations with an emphasis on text annotations~\cite{ hullman2013contextifier} (see \autoref{sec.prev-design-space} for more details). However, there is a lack of comprehensive analysis covering diverse annotation types, encompassing both textual and graphical forms, their functions, and applicability across a wide range of chart types. This gap highlights the need for a design space that articulates the various annotation options suited to different chart types. 

In this paper, we present a qualitative study to understand the design space for annotating visualizations, focusing on the common practices of annotations from diverse real-world scenarios across a wide range of chart types. To achieve this understanding, we evaluated over 1,800 static chart images with annotations (e.g., annotated images of bar charts, line charts, scatterplots, etc.) collected from Google Images. \autoref{fig:teaser} illustrates several examples of the annotated charts we obtained. Through thematic analysis~\cite{braun2012thematic}, we identified recurring patterns in the usage of annotations, which allowed us to construct a design space of annotations grounded in common practices.

We situated our design space of annotations within Brehmer and Munzner's task typology~\cite{brehmer2013multi} (see~\autoref{sec.design_space}), employing its \textit{"Why? How? What?"} framework for a systematic and structured approach to analyze and construct annotations in visualizations. The \taskWhy reveals the tasks from the typology for which people use annotations, such as \texttt{presenting}, \texttt{identifying}, \texttt{summarizing}, and \texttt{comparing} visual elements. To characterize \taskHow, we identified a taxonomy of seven annotation types, including \textit{enclosures, connectors, text, glyphs, color, indicators}, and \textit{geometric}, and explored their common uses across chart types. We also discuss the concept of annotation ensembles for situations where a single type fails to convey the message clearly, demonstrating how combining annotations, such as connectors and text, can enhance clarity. The \taskWhat dimension categorized annotations into three categories based on their relationship with the data sources that generated them: directly from the data itself, derived from the data, or external knowledge brought by the designer. 

We discuss the utility of our proposed design space through three case studies, demonstrating its practical application in enhancing visualizations through targeted annotations based on specific tasks and common practices (see~\autoref{sec.utility}). Our design space provides a systematic, actionable framework for visualization practitioners, professionals, and researchers to select annotations from a range of options and annotate diverse chart types tailored to their viewers' needs.
In summary, the contributions of this research are: 

\begin{itemize}[noitemsep,itemsep=4pt]
    \item We analyzed annotation usage patterns from N=1,888 annotated charts across 14 chart types to construct a taxonomy of seven distinct annotation types.

    \item  We proposed a design space of annotations in an actionable \textit{"Why? How? What?"} framework, where \taskWhy discusses the tasks annotations support, \taskHow examines methods and combinations based on common practices, and \taskWhat identifies necessary data sources for annotation, thereby offering a structured approach to annotating visualizations.

    \item We have provided a dataset of annotated charts of different types, offering valuable real-world examples of annotation practices for researchers and practitioners to leverage in future studies. Our dataset is available at \url{https://shorturl.at/bAGM1}.

\end{itemize}

\section{What is an Annotation?}
\label{sec.taxonomy-and-design-space}

Before building a design space of annotations, it was important to identify a clear definition of what annotations are in our context.

Our analysis of annotations in visualization literature revealed a variety of definitions, each emphasizing different aspects. Munzner and Chen et al.\ defined annotations as enhancements in the form of additional graphical or textual elements that serve to introduce new data attributes and provide additional context, thereby clarifying the visualization's information to the viewers~\cite{munzner2014visualization, chen2023does}. Kong et al.\ viewed annotations as a graphical overlay designed to enable communication and collaborative analysis of visualizations while minimizing visual clutter~\cite{kong2012graphical}. In their study, Kong et al.\ defined annotation as a visual element added or modified to guide viewers' attention to a specific area of a visualization~\cite{kong2017internal}. Given what we observed in prior works, it was clear that the specifics of annotations were subjective. Therefore, although our definition aligns with the prior definitions, we used the following definition of annotations for clarity:

\noindent
\vspace{-5pt}
\begin{center}
\begin{minipage}{0.93\linewidth}
\textit{Annotations are supplementary elements, such as graphical shapes, text, or color, added to pre-existing elements of a visualization to provide additional context beyond the basic data presented and to draw attention to specific portions or elements of the data, thereby enhancing the expressiveness of the visualization.}
\end{minipage}
\end{center}
\vspace{5pt}

The basic elements of a chart encompass constituent elements necessary for conveying the relevant information without reliance on external aids (usually but not exclusively axes, axis labels, data points, titles, legends, and gridlines). Conversely, the supplementary elements of a chart are intended to augment the comprehensibility and expressiveness of the chart (additional textual or graphical elements). For instance, \autoref{fig:cs3-a} demonstrates a line chart without annotations, and \autoref{fig:cs3-d} shows the exact visualization with annotations added. The process of adding these annotations (\autoref{fig:cs3-b} and \ref{fig:cs3-c}) will be described in \autoref{sec.utility:cs3}.

\subsection{Design Spaces of Annotations}
\label{sec.prev-design-space}

Research in the existing visualization literature has examined annotation design spaces with a limited focus on certain annotation types, chart formats, or application domains. For instance, Ren et al.\ investigated professional charts designed for visual data storytelling and developed a design space of annotations categorized by their form and target~\cite{ren2017chartaccent}. They outlined four forms of annotations: text, shapes, highlights, and images---each serving distinct purposes---text annotations for adding narrative or insights, shape annotations for drawing focus through geometric forms, highlights for visual emphasis or de-emphasis of elements, and image annotations for additional graphical context. Their analysis further identified different annotation targets, such as data points, series, chart elements, and coordinate spaces. Although their research explored both textual and graphical annotations, their analysis was limited to exclusively bar charts, line charts, and scatterplots explicitly designed for visual storytelling.

Furthermore, Hullman et al.\ analyzed professional narrative visualizations where they found two main categories of annotations---additive, which incorporates external information to provide broader context or background, and observational, which focuses on the data presented by drawing attention to specific values, trends, or outliers within the visualization itself~\cite{hullman2013contextifier}. The study also explored levels of anchoring for annotations: anchoring to a single datum, anchoring to a group/region, and anchoring to the entire visualization. This study looked at a comparatively broad set of seven common chart types, but they focused exclusively on textual annotations in narrative storytelling.

Our research extends beyond these studies by examining the use of both textual and graphical annotations across a diverse set of 14 common chart types collected from Google Image search. We introduce a design space based on an actionable framework that categorizes annotation types, outlines their functional roles, highlights common uses, and discusses their data sources, making it relevant to a broader spectrum of data visualization applications.

\section{The Roles of Annotations in Visualizations}
\label{section:background}

In this section, we explore the multifaceted roles of annotations in enhancing data visualizations.

\paragraph{Comprehension, Memorability, and Recall} 
Prior research has highlighted the critical role of annotations in enhancing comprehension, memorability, and recall within visualizations. Studies have shown that annotations enhance long-term recall and deepen viewers' understanding and information retention~\cite{bateman2010useful, zheng2021sketchnote}. Eye-tracking studies have emphasized the critical role of annotations in directing attention and facilitating memory encoding, underscoring their importance in creating memorable visualizations~\cite{borkin2013makes, borkin2015beyond}. Further analysis has revealed the varied effectiveness of annotations: additive annotations are shown to improve comprehension and recall across contexts~\cite{chun2020giving}, whereas the effectiveness of observational annotations and specific visual cues in narrated visualizations varies, highlighting a complex interplay of effects~\cite{kong2019understanding}.

\paragraph{Narrative Storytelling} 
Annotations significantly enhance data visualization storytelling by improving comprehension, engagement, and clarity, ensuring coherence; and facilitating interactive exploration in complex narratives without overwhelming the audience~\cite{segel2010narrative, hullman2013deeper, kosara2013storytelling, boy2015storytelling}. The integration of visuals and text is shown to significantly enhance comprehension, with detailed annotations in charts preferred by users for better interpretation and reducing bias perceptions~\cite{ottley2019curious, stokes2022striking, stokes2023role, fan2024understanding}. By highlighting key features, suggesting conclusions, and providing context, textual annotations not only enrich the narrative quality but also make complex data more accessible and engaging by directing user interpretations~\cite{hullman2011visualization, martinez2020data, walker2015storyboarding, bryan2016temporal, stolper2016emerging, rahman2023exploring}. Moreover, in multimedia formats such as data comics and videos, annotations play a crucial role in strengthening narrative cohesion and augmenting viewers' understanding~\cite{wang2019comparing, shi2021communicating, zhao2015data, bach2018design}. The development of automated annotation techniques, combined with manual efforts, amplifies narrative capabilities and interactive storytelling. Tools such as SketchStory~\cite{lee2013sketchstory}, Ellipsis~\cite{satyanarayan2014authoring}, Narvis~\cite{wang2018narvis}, Charagraphs~\cite{masson2023charagraph}, Timeline Storyteller~\cite{brehmer2019timeline}, and NewsViews~\cite{gao2014newsviews} underscore the dynamic role of annotations in deepening understanding and enriching storytelling in data visualization.

\paragraph{Collaborative Sensemaking} Research shows that annotations are crucial for collaborative sensemaking in visualizations by facilitating the interpretation of complex datasets, bridging communication gaps, and enhancing collective understanding~\cite{sanderson1994exploratory, robinson2008collaborative, chen2011supporting, kadivar2009capturing}. They are instrumental in various contexts, from organizational knowledge sharing to co-located settings, improving group performance and sensemaking~\cite{bresciani2009benefits, mahyar2012note}. Platforms such as Many Eyes~\cite{viegas2007manyeyes} and communication-minded visualization (CMV)~\cite{viegas2006communication} illustrate how annotations support community engagement and understanding, with tools such as sense.us~\cite{heer2007voyagers} and Click2Annotate~\cite{chen2010click2annotate} showing their importance in asynchronous collaborations~\cite{heer2007design}. Further studies highlight the role of annotations in knowledge handoff and collaborative interpretation alongside systems that foster teamwork and effective communication through interactive exploration, underscoring the broad applicability of annotations in collaborative visual analytics~\cite{kong2009perceptual, ellis2004collaborative, eccles2008stories, schwab2020visconnect, mathisen2019insideinsights}.

\paragraph{Other Applications}
Apart from the applications above, annotations are instrumental in enhancing user interaction and engagement~\cite{kong2017internal, kauer2021public, wood2012sketchy, taher2015exploring}, helping externalize and explore data~\cite{Kang2014characterizing, sevastjanova2021visinreport, mahyar2012note, shrinivasan2009connecting, kim2019inking, choe2015characterizing, heer2012interactive, lin2022data, romat2019activeink}, playing a vital role in provenance visualizations~\cite{gadhave2022reusing, groth2006provenance, gadhave2021predicting, gratzl2016visual, stitz2018knowledgepearls, lin2021sanguine}, and assisting in visualizing uncertainty~\cite{cedilnik2000procedural, liu2016uncertainty, ferreira2014sample}, and they are key in supporting visual debugging~\cite{hopkins2020visualint, wood2012sketchy, hoffswell2016visual, Arlene2022annotating}, underscoring their widespread relevance across different visualization domains.

\section{Real-World Annotation Practices}

We investigated how annotations were applied to diverse chart types in real-world settings. To accomplish this, we obtained a large corpus of annotated images featuring commonly used chart types from Google Images and subjected them to a thematic coding procedure to ascertain patterns in the application of annotations.

\begin{figure}[!bp]
    \centering
    \includegraphics[width=0.9\linewidth]{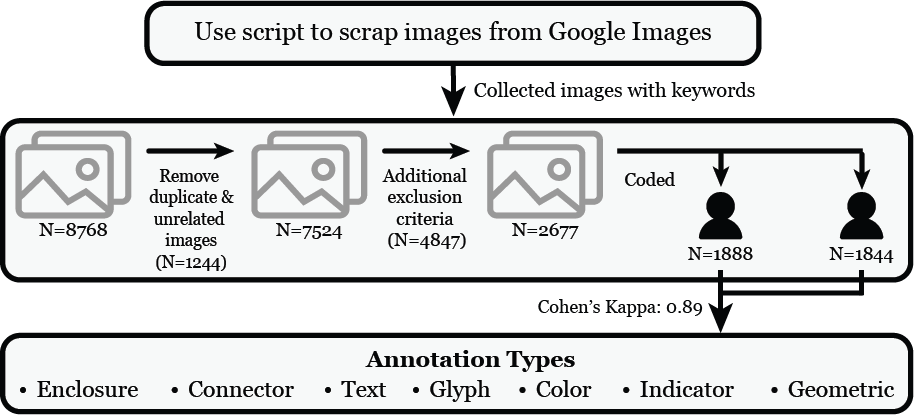}
    
    \caption{We scrapped annotated visualization images from Google Images using the search pattern ``\textit{annotated \{chart type\}}'', where the chart type included 14 commonly used charts (e.g., scatterplot). Two coders qualitatively coded the images into the seven annotation types shown.}
    \label{fig:coding}
\end{figure}

\subsection{Data Collected}
Our goal was to collect a large number of charts from diverse applications. Therefore, we obtained 8,768 images from Google Images using an image scrapping tool~\cite{ohyicong} (see \autoref{fig:coding}). Recognizing the challenge of comprehensively creating search queries covering a wide range of real-world scenarios, we kept our queries as straightforward as possible. We used different query keywords following the same pattern, ``\textit{annotated \{chart type\}}'' (e.g., annotated pie chart, annotated scatterplot, annotated histogram, etc.). The queries used many common chart types, including \textit{line chart}, \textit{bar chart}, \textit{map}, \textit{scatterplot}, \textit{pie chart}, \textit{bubble chart}, \textit{donut chart}, \textit{area chart}, \textit{treemap}, \textit{histogram}, \textit{graph} (i.e., node-link diagrams), \textit{gantt chart}, \textit{density map}, and \textit{radar chart}. From the collected corpus of charts, we observed that the distribution among the chart types was not uniform, likely indicative of their popular use.

We removed 1,244 duplicate images from the dataset during our initial evaluations of the images. 
We then removed 4,847 based upon additional exclusion criteria, including images with chart types not considered in the study, images with charts that are not clearly visible, images with charts without annotations, and images with charts that do not represent real data. This left a total of 2,677 images in the dataset.

\subsection{Data Coding}

We employed a thematic coding process to analyze our dataset. The themes were based on five annotation types from a prior study~\cite{rahman2022qualitative}, in which visualization students annotated bar charts to answer high-level questions through specific low-level tasks~\cite{amarstasko}. This study led to a taxonomy of bar chart annotations, namely \textit{enclosures}, \textit{connectors}, \textit{text}, \textit{glyphs}, and \textit{color}, which we then used as initial themes in our coding process. Two coauthors independently coded all 2,677 images in multiple iterations with frequent meetings involving all authors to discuss and update the themes. Ultimately, the two coders identified 1,888 and 1,844 images, respectively, where the chart types relevant to the study were present, and the annotation type was not classified as ``no annotation'', ``undetermined'' (i.e., ambiguous), or ``other'' (i.e., not a commonly practiced annotation type). We evaluated the inter-rater reliability of our coding process using the Kappa statistic~\cite{mchugh2012interrater}. We compared the annotation types for the corresponding images between the raters and calculated the average Cohen's Kappa coefficient of 0.886, which indicated a very high agreement between the raters. Thus, for data analysis, we utilized the annotations from the first coder. Supplementary materials provide the coding from both coders.

\subsection{Taxonomy of Annotation Types}

\begin{figure}[!t]
    \centering

    \begin{minipage}[b]{0.95\linewidth}
        {\includegraphics[width=0.975\linewidth]{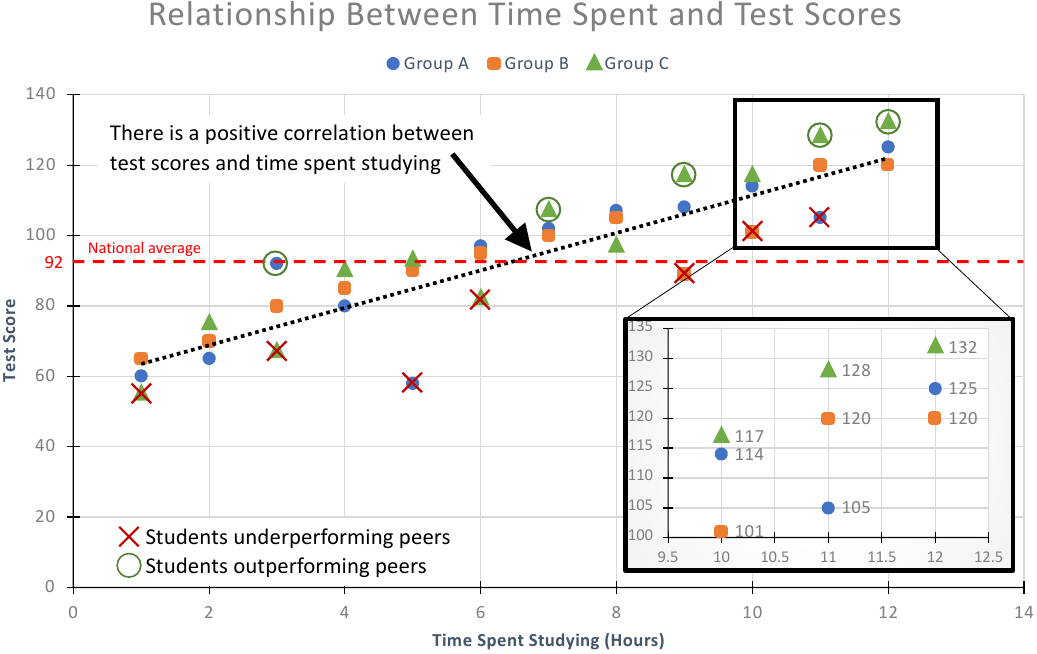}}
    \end{minipage}
    \hspace{-1pt}
    \begin{minipage}[b]{1pt}
        \hspace{-27pt}
        \begin{minipage}[t][8pt][c]{6pt}\drawCircle{black}{LightGray}{5pt}\end{minipage}

        \vspace{-12pt}
        \hspace{-24.5pt}
        \subfloat[\label{fig:illustrative-scatterplot:a}]{}
        \vspace{105pt}
    \end{minipage}%
    \begin{minipage}[b]{0pt}
        \hspace{-146pt}
        \begin{minipage}[t][8pt][c]{6pt}\drawCircle{black}{LightGray}{5pt}\end{minipage}

        \vspace{-12pt}
        \hspace{-143.5pt}
        \subfloat[\label{fig:illustrative-scatterplot:b}]{}
        \vspace{85pt}
    \end{minipage}%
    \begin{minipage}[b]{0pt}
        \hspace{-225pt}
        \begin{minipage}[t][8pt][c]{6pt}\drawCircle{black}{LightGray}{5pt}\end{minipage}

        \vspace{-12pt}
        \hspace{-222.5pt}
        \subfloat[\label{fig:illustrative-scatterplot:c}]{}
        \vspace{110pt}
    \end{minipage}%
    \begin{minipage}[b]{0pt}
        \hspace{-217.5pt}
        \begin{minipage}[t][8pt][c]{6pt}\drawCircle{black}{LightGray}{5pt}\end{minipage}

        \vspace{-12pt}
        \hspace{-215pt}
        \subfloat[\label{fig:illustrative-scatterplot:d}]{}
        \vspace{20pt}
    \end{minipage}%
    \begin{minipage}[b]{0pt}
        \hspace{-204pt}
        \begin{minipage}[t][8pt][c]{6pt}\drawCircle{black}{LightGray}{5pt}\end{minipage}

        \vspace{-12pt}
        \hspace{-201.5pt}
        \subfloat[\label{fig:illustrative-scatterplot:e}]{}
        \vspace{49pt}
    \end{minipage}%
    \begin{minipage}[b]{0pt}
        \hspace{-220pt}
        \begin{minipage}[t][8pt][c]{6pt}\drawCircle{black}{LightGray}{5pt}\end{minipage}

        \vspace{-12pt}
        \hspace{-217pt}
        \subfloat[\label{fig:illustrative-scatterplot:f}]{}
        \vspace{69pt}
    \end{minipage}%
    \begin{minipage}[b]{0pt}
        \hspace{-111.5pt}
        \begin{minipage}[t][8pt][c]{6pt}\drawCircle{black}{LightGray}{5pt}\end{minipage}

        \vspace{-12pt}
        \hspace{-109pt}
        \subfloat[\label{fig:illustrative-scatterplot:g}]{}
        \vspace{60pt}
    \end{minipage}

    \caption{An illustrative example of annotations in a scatterplot showing (a)~\textit{enclosure} of a group of data, (b)~a \textit{connector} between text and an indicator, (c)~\textit{text} describing the result, (d)~\textit{glyph} and \textit{color} that highlight certain data points, two \textit{indicator} annotations show (e)~correlation and (f)~average, and (g)~a \textit{geometric} annotation in the form of a zoom box.}
    \label{fig:illustrative-scatterplot}
\end{figure}

Upon analysis of our data, we identified seven annotation types applicable to various chart types: \textit{enclosures}, \textit{connectors}, \textit{text}, \textit{glyph}, \textit{color}, \textit{indicator}, and \textit{geometric} annotations. \autoref{fig:illustrative-scatterplot} provides examples of each of these annotation types.

\vspace{3pt}
\begin{wrapfigure}[3]{l}{0.075\linewidth}
    \vspace{-12pt}
    {\includegraphics[height=1.0cm]{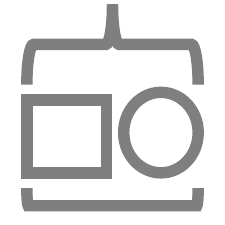}}
\end{wrapfigure}
\noindent\paragraph{Enclosure}
Enclosure is an annotation that uses a partially or fully closed boundary, including ellipses, brackets, rectangles, etc. These enclosure annotations find application in a variety of situations. For instance, in \autoref{fig:illustrative-scatterplot:a}, a rectangle identifies a group of data points within the scatterplot, and in \autoref{fig:teaser:d_old_b}, a rectangle is employed for a similar purpose within the node-link diagram. Additionally, in \autoref{fig:teaser:a}, ellipses are used with connectors (i.e., arrows) to emphasize specific numbers in the bar chart.

\vspace{3pt}
\begin{wrapfigure}[3]{l}{0.075\linewidth}
    \vspace{-12pt}
    {\includegraphics[height=1cm]{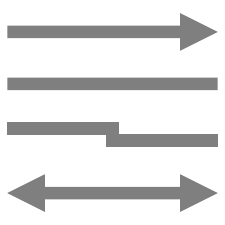}}
\end{wrapfigure}
\noindent\paragraph{Connector}
A connector annotation involves using a line encompassing various forms, such as solid, dotted, or directional (i.e., arrows) lines. For instance, in \autoref{fig:illustrative-scatterplot:b}, the connector points the text to the element it refers to. The role of arrows is exemplified in \autoref{fig:teaser:a}, where they are employed with ellipses to visually represent distinctions between different years. Furthermore, in \autoref{fig:teaser:c}, lines are employed to establish connections between specific points of interest on the line and corresponding text descriptions. Similarly, in \autoref{fig:teaser:e}, lines and text pair to highlight significant events on the map.

\vspace{3pt}
\begin{wrapfigure}[3]{l}{0.075\linewidth}
    \vspace{-12pt}
    {\includegraphics[height=1.0cm]{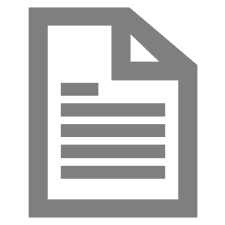}}
\end{wrapfigure}
\noindent\paragraph{Text}
Text annotation involves using words or sentences to provide answers regarding the data. Text primarily functions as descriptions, explaining phenomena in the data and events depicted in charts, or indicating specific dataset values. For instance, in \autoref{fig:teaser:a} and \autoref{fig:teaser:c}, text values pinpoint significant aspects within the charts. In \autoref{fig:teaser:c} and \autoref{fig:teaser:e}, textual descriptions elaborate on different events shown on the line chart and the map, respectively. Additionally, in \autoref{fig:illustrative-scatterplot:c}, text descriptions, along with arrows, clarify the positive correlation between two scatterplot variables.

\vspace{3pt}
\begin{wrapfigure}[3]{l}{0.075\linewidth}
    \vspace{-12pt}
    {\includegraphics[height=1.0cm]{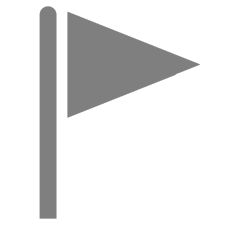}}
\end{wrapfigure}
\noindent\paragraph{Glyph}
Glyphs are annotations that employ symbols or shapes to address data-related queries and identify specific objects or categories. For instance, in \autoref{fig:teaser:c}, circular glyphs draw attention to the focal point on the line in the line chart. In the illustrative figure shown in \autoref{fig:illustrative-scatterplot:d}, glyphs, combined with colors, distinguish data points of various types in the scatterplot.

\vspace{3pt}
\begin{wrapfigure}[3]{l}{0.075\linewidth}
    \vspace{-12pt}
    {\includegraphics[height=1.0cm]{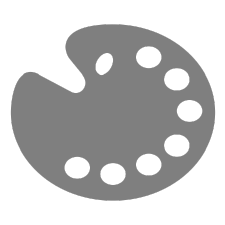}}
\end{wrapfigure}
\noindent\paragraph{Color}
Color annotations are any property of color (generally, but not exclusively, hue) to answer questions about data. Colors commonly serve to spotlight specific segments of a chart or to distinguish between various categories. For instance, in \autoref{fig:teaser:f}, color effectively sets apart and emphasizes the lacrosse slice from other sections in the pie chart. Similarly, in \autoref{fig:illustrative-scatterplot:d}, distinct colors differentiate various glyphs, even when they possess differing shapes (e.g., cross and circular), underscoring their disparities.

\vspace{3pt}
\begin{wrapfigure}[3]{l}{0.075\linewidth}
    \vspace{-12pt}
    {\includegraphics[height=1.0cm]{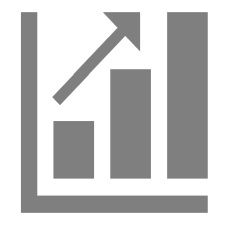}}
\end{wrapfigure}
\noindent\paragraph{Indicator}
For several chart types, we found many annotations that indicate trends, statistics, thresholds, etc.\ in the data. For example, we can see a line in \autoref{fig:illustrative-scatterplot:e} that shows that there is a positive correlation between \textit{test scores} of students from different groups and \textit{time spent studying}. \autoref{fig:cs1-c} also uses lines to indicate declining trends of deaths of despairs.

\vspace{3pt}
\begin{wrapfigure}[3]{l}{0.075\linewidth}
    \vspace{-12pt}
    {\includegraphics[height=1.0cm]{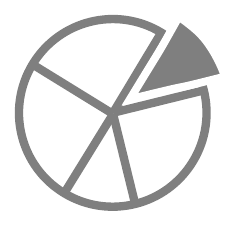}}
\end{wrapfigure}
\noindent\paragraph{Geometric}
Geometric annotation involves modifying chart areas to highlight specific information, such as enlarging a specific area or portion of a chart and zooming in a particular part (see \autoref{fig:illustrative-scatterplot:g}). Examples include enlarging a specific pie segment to identify a particular piece of information (see \autoref{fig:teaser:f}) and zooming in on a particular part of a graph (see \autoref{fig:teaser:d_old_b}).

\section{Design Space for Annotations}
\label{sec.design_space}

\begin{figure*}[!t]
    \centering
    
    \includegraphics[width=0.97\linewidth]{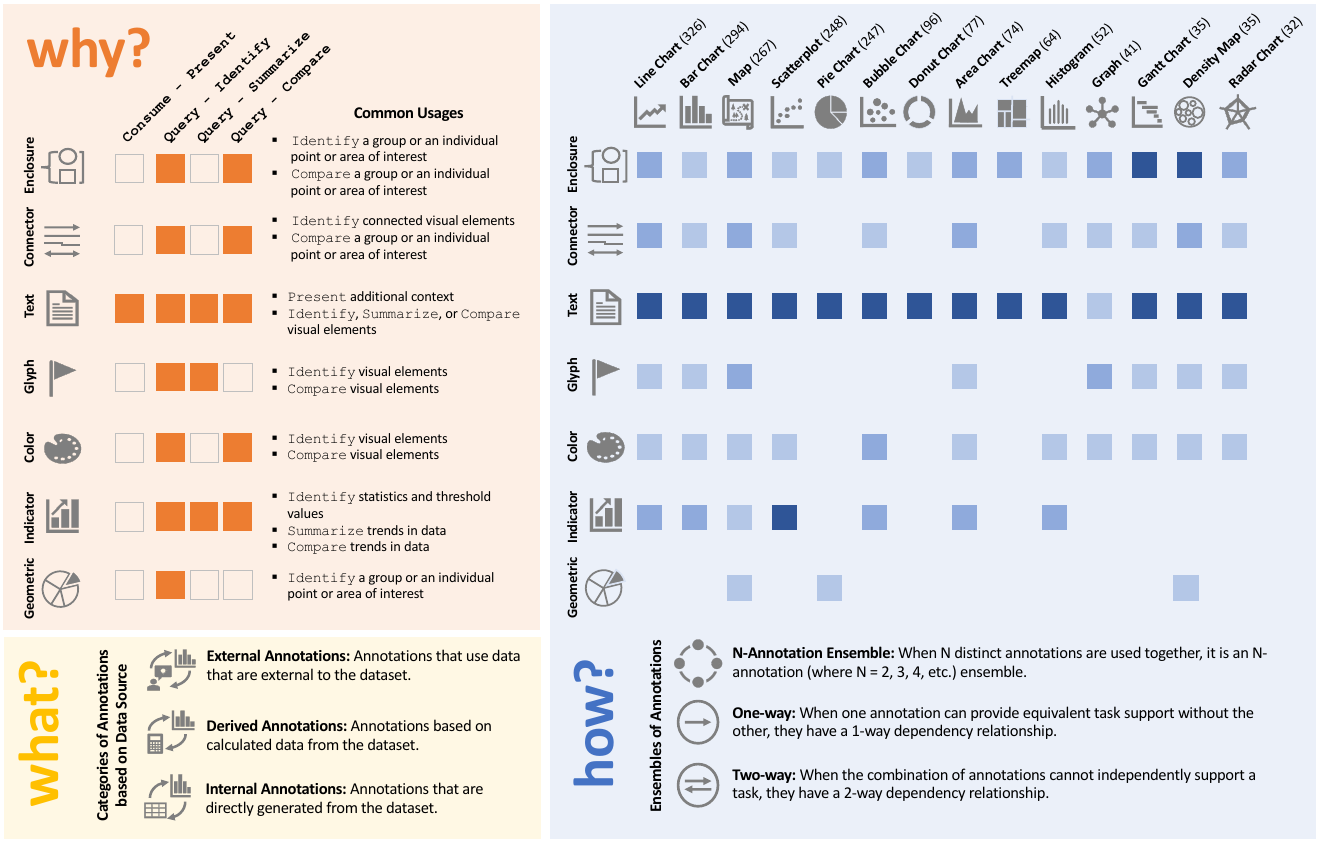}
    
    \caption{Our design space of annotations is divided into three key sections. The design space is used by starting in the \taskWhy section, which identifies a task and potential annotation types. Then, \taskHow elaborates common usages of annotation in two parts: a color-coding system indicating the usage frequency of annotations (\fcolorbox{black}{dsLow}{\tiny 6-25\%}, \fcolorbox{black}{dsMed}{\tiny 26-50\%}, and \fcolorbox{black}{dsHigh}{\tiny \textcolor{white}{51+\%}}), and the types of annotation ensembles. Finally, \taskWhat is used to categorize the annotation data source.
    }
    \label{fig:design-space}

\end{figure*}

Annotations, like other aspects of visualizations, are deliberately designed to accomplish specific communicational tasks or goals, serving as a critical component of visualizations. Therefore, we found it useful to situate the design space of annotations within a well-established visual analytics task framework, particularly within the \textit{``Why? How? What?''} framework of Brehmer and Munzner~\cite{brehmer2013multi}. 

We addressed three main questions derived from their task typology, specifically focusing on the process of using annotations in charts within our design space:

\vspace{3pt}\noindent
\taskWhy --- aims to understand the analytic purposes of using annotations in charts (\autoref{sec.why}). It helps designers identify the tasks they want to support with annotations.

\vspace{3pt}\noindent
\taskHow --- explores the strategies available to annotate a chart (\autoref{sec.how}). It assists designers in navigating the variety of annotation options available, tailored to their analytic purposes.

\vspace{3pt}\noindent
\taskWhat --- explores the types of data needed to generate the chosen annotations (\autoref{sec.what}). It assists designers in identifying the relationship between the annotations and the data from which they would be created.

The interrelation among these three questions provides a structured and systematic framework for analyzing and constructing annotations in visualizations aimed at improving how data is communicated and understood.

\vspace{2pt}
\subsection{\taskWhy\ --- Tasks for Annotations} 
\label{sec.why}
\vspace{-3pt}
There are multiple reasons for annotating a chart, including offering context via textual explanations or emphasizing particular data elements through methods such as zooming or highlighting critical data points. Exploring the analytical purpose behind using annotations in visualizations is the focus of \taskWhy. Although it was challenging to analyze the narrative goals in our dataset without direct communication with the visualization designers, our examination identified several clear tasks for annotation usage, e.g., textual annotations in~\autoref{fig:teaser:a} that help the viewers \texttt{compare} bars; enclosures in~\autoref{fig:teaser:b__old_d} that help the viewers \texttt{identify} clusters within scatterplots; textual annotations in~\autoref{fig:teaser:c} that help designers \texttt{present} external events over time with connectors for precise temporal association; and indicators in~\autoref{fig:illustrative-scatterplot:e} that help the viewers \texttt{summarize} trends in data. For visualization designers, \taskWhy provides a framework to guide the selection of annotation types that align with their specific communication goals, leading to annotations that support the intended tasks of the viewers.

In particular, we observed that the majority of annotations supported \textit{querying} tasks that designers are directing their viewers to perform. Specifically, all annotation types could be used for \texttt{identify} tasks, most could be used for \texttt{compare} tasks, and text, glyphs, and indicators provided the capability to \texttt{summarize} (see~\cite{brehmer2013multi} for definitions of these tasks). Finally, only text provided the ability to \texttt{present} additional context to the visualization, which Brehmer and Munzner defined as a \textit{consume} task. These insights facilitate a deeper understanding of the various motivations behind employing annotations, systematically encapsulated in the \taskWhy section of our proposed design space, referenced in \autoref{fig:design-space}.

\subsection{\taskHow\ --- Patterns of Annotation Usage}
\label{sec.how}
\vspace{-2pt}

We address the \taskHow question by detailing the usage of different annotation types (\autoref{sec.types}) and their ensembles (\autoref{sec.ensemble}). This explanation extends to their application across various chart types for tasks outlined in the \taskWhy section, emphasizing the strategic selection and implementation of annotations to meet communication goals.

\subsubsection{Common Patterns for Annotation Types}
\label{sec.types}

We examine how different annotation types are used in different chart types so that practitioners and designers can gain insights into common practices of annotation usage in visualization design. Color-coded boxes in \taskHow in~\autoref{fig:design-space} show the percentage (\%) of different annotation types used in various charts.

\vspace{3pt}
\begin{wrapfigure}[3]{l}{0.075\linewidth}
    \vspace{-12pt}
    {\includegraphics[height=1.0cm]{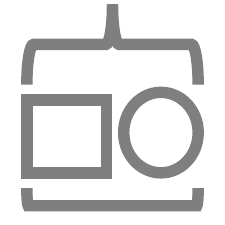}}
\end{wrapfigure}
\noindent\paragraph{Enclosures} Enclosure annotations were frequently utilized in all chart types, appearing in 588 of 1,888 charts in our dataset ($\sim$31\%). These annotations were primarily used to \texttt{identify} or \texttt{compare} a group or an individual point or area of interest to make it easier for viewers to interpret the data. For example, in~\autoref{fig:illustrative-scatterplot:a}, a rectangle has been used to identify a group of data points. Enclosures can also be used in various other ways as an identifier, such as highlighting a particular bar in a bar chart, a specific section of a line chart, a particular area of interest in maps, or particular nodes in a graph (see~\autoref{fig:teaser:d_old_b}). Furthermore, enclosures can act as a container (i.e., identify) for other annotations (see~\autoref{fig:teaser:a}) and separate (i.e., compare) the enclosed data from other elements within the chart.

\vspace{3pt}
\begin{wrapfigure}[3]{l}{0.075\linewidth}
    \vspace{-12pt}
    {\includegraphics[height=1.0cm]{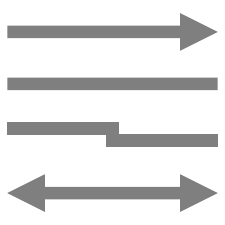}}
\end{wrapfigure}
\noindent\paragraph{Connectors} We observed connectors, such as arrows and lines, in most chart types appearing in 393 charts ($\sim$21\%). However, they were primarily used in conjunction with other annotation types (more discussion in~\autoref{sec.ensemble}). Their purpose is to \texttt{identify} visual connections between elements within charts. For example, an arrow connects the point of interest and the text description or enclosure containing text description, making it easier for viewers to understand which annotation corresponds to which point of interest in~\autoref{fig:illustrative-scatterplot:b}. Connectors can also be used to \texttt{compare} different data points or areas of interest within a chart, such as connecting two bars in a bar chart to show the difference between them as shown in~\autoref{fig:teaser:a}. We also observed connectors customized with color, thickness, and other style parameters, allowing for flexible usage and visual impact.

\vspace{3pt}
\begin{wrapfigure}[3]{l}{0.075\linewidth}
    \vspace{-3pt}
    {\includegraphics[height=1.0cm]{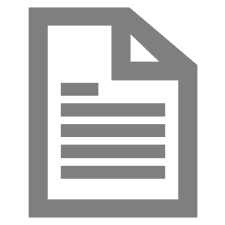}}
\end{wrapfigure}
\noindent\paragraph{Text}
Our analysis revealed a prevalent use of text annotations across all chart types, with text annotations appearing in 1,434 out of 1,888 charts ($\sim$76\%), underscoring a clear preference for them over other annotation types in practice. The widespread use of text annotations across all chart types is attributable to their flexibility, which enables conveying a diverse range of information. Text annotations can be employed to provide simple labels, detailed descriptions, brief summaries, or in-depth explanations, depending on the purpose of the visualization and the viewers' needs. Text annotations were utilized in diverse ways across different chart types, serving a range of purposes, such as to \texttt{present} additional context and to \texttt{identify}, \texttt{compare}, and \texttt{summarize} visual elements within charts. For instance, in~\autoref{fig:illustrative-scatterplot:c}, despite the presence of a correlation line to denote a positive correlation between the two variables, a text description combined with an arrow (i.e., connector) is used to present additional context about the relationship. In~\autoref{fig:teaser:a}, text values inside the enclosures help viewers compare the bars, and in~\autoref{fig:teaser:c}, text descriptions summarize information from the related article and help them identify the points of interest on the line chart.

\vspace{3pt}
\begin{wrapfigure}[3]{l}{0.075\linewidth}
    \vspace{-12pt}
    {\includegraphics[height=1.0cm]{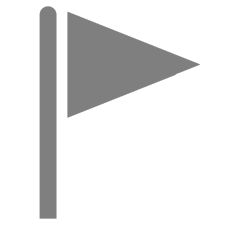}}
\end{wrapfigure}
\noindent\paragraph{Glyphs}
Glyphs were mainly used to \texttt{identify} or \texttt{compare} visual elements in charts, appearing in 270 charts ($\sim$14\%). We noticed the most prominent use of glyphs in maps. In maps, glyphs were used to identify different points of interest or regions. Also, glyphs were used in line charts to identify specific points of interest along the plotted line, as depicted in~\autoref{fig:teaser:c}. Similarly, glyphs were applied in bar charts to identify the bar(s) of interest or specific point(s) on any chart axes. In scatterplots, glyphs were employed to compare particular data points from others or to highlight individual or multiple points of interest. For example, in~\autoref{fig:illustrative-scatterplot:d}, glyphs are employed with color to identify and compare the data points representing students who are underperforming and outperforming their peers.

\vspace{3pt}
\begin{wrapfigure}[3]{l}{0.075\linewidth}
    \vspace{-12pt}
    {\includegraphics[height=1.0cm]{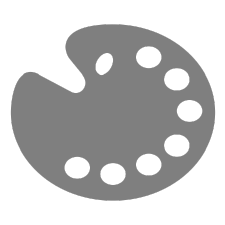}}
\end{wrapfigure}
\noindent\paragraph{Color}
Color effectively distinguishes between different categories or data points and draws attention to specific elements or regions of a visualization, appearing in 259 charts ($\sim$14\%). 
The primary function of color is to \texttt{identify} or \texttt{compare} visual elements by highlighting a specific data point or set of data points. For instance, in~\autoref{fig:illustrative-scatterplot:d}, color has been used with glyphs to identify the difference between some data points from others. Here, data points representing underperforming students are denoted by red-colored cross glyphs, whereas green-colored ellipses indicate data points representing outperforming students.
Additionally, color is employed to highlight specific regions of a chart, such as slices of interest in pie charts (see~\autoref{fig:teaser:f}). It is important to note that encodings inherent to the charts are not considered annotations, such as color encoding for different data points indicating different categories in a scatterplot or color encoding for different bars indicating different categories in a grouped bar chart using legends.

\vspace{3pt}
\begin{wrapfigure}[3]{l}{0.075\linewidth}
    \vspace{-12pt}
    {\includegraphics[height=1.0cm]{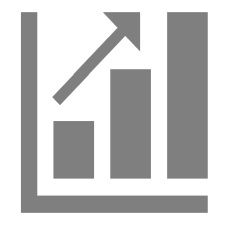}}
\end{wrapfigure}
\noindent\paragraph{Indicator} 
Indicators appearing in 424 charts ($\sim$20\%) are prominently used in line charts, bar charts, scatterplots, bubble charts, histograms, and area charts. In bar charts, histograms, line charts, and area charts, indicators were utilized to \texttt{identify} or \texttt{compare} changes in values along a particular axis, whether in direction (e.g., upward or downward trend) or magnitude (e.g., rapid or slower change). Conversely, in scatterplots (see~\autoref{fig:illustrative-scatterplot:e}) and bubble charts, trend lines were primarily used as indicators to \texttt{summarize} the correlation relationship (e.g., positive or negative correlation) between variables, along with the strength of the correlation relationship (e.g., strong or weak correlation). Furthermore, indicators are also used to identify thresholds, statistical values, or benchmark points along a specific axis in certain chart types (see~\autoref{fig:illustrative-scatterplot:f}). However, such indicators are uncommon in some chart types, primarily due to the absence of a clear axis. For instance, pie charts represent proportions, and the slices do not have any inherent order. Similarly, treemaps do not have an explicit ordering of the data and no linear scale.

\vspace{3pt}
\begin{wrapfigure}[3]{l}{0.075\linewidth}
    \vspace{-12pt}
    {\includegraphics[height=1.0cm]{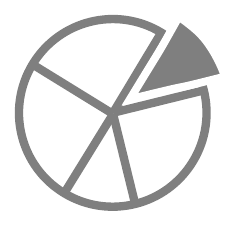}}
\end{wrapfigure}
\noindent\paragraph{Geometric} Geometric annotations, appearing in 61 charts ($\sim$3\%), were more prevalent in certain chart types, such as pie charts, donut charts, and maps, to \texttt{identify} a group or an individual point of interest to draw viewers' attention. For example, in pie and donut charts, where each wedge or slice represents a portion of a whole, it may be necessary to emphasize (i.e., identify) a particular section if it contains crucial information or if it is of significant interest to the viewer (see~\autoref{fig:teaser:f}). In \autoref{fig:illustrative-scatterplot:g}, the scatterplot has been zoomed in on a specific portion to provide additional context to the viewers. Similarly, in maps, zooming in on a specific area focuses attention on important regions while providing greater details about those areas. 
The suitability of geometric annotations relies on the chart type, the data type, and the message the visualization intends to convey. For example, zooming in on a particular bar or section may not be necessary or effective in a bar chart. Similarly, in a line chart, zooming in on a specific section may cause the viewer to lose sight of the overall trend in the data.

\subsubsection{Common Patterns for Ensemble Usage}
\label{sec.ensemble}

We observed that in cases where a single annotation type was insufficient to convey the intended information, people resorted to using a combination of multiple annotations, a practice we refer to as the ensemble of annotations. Our investigation has uncovered significant use of ensemble annotations in the charts we analyzed. Although a definitive enumeration of the number of ensemble annotations was not conducted due to the lack of explicit information regarding the corresponding annotation designer intention, we observed consistent patterns of ensemble usage in our analysis. The most frequently used ensembles across different chart types were enclosure+connector+text, connector+text, enclosure+text, and glyph+color. In these ensembles, text annotations were primarily used to provide additional context, enclosures were utilized to identify or separate text annotations, whereas connectors were used to identify the connection between the corresponding data point within the chart and the text or the enclosure with text. For instance, in~\autoref{fig:illustrative-scatterplot:b}, a connector (i.e., the arrow) is used to identify the connection between the correlation line and the text, presenting the correlative relationship between the two variables in the scatterplot. Color was also frequently used with glyphs, text, connectors, and enclosures to identify or compare specific categories in different charts. For example, in~\autoref{fig:illustrative-scatterplot:d}, color is used with glyphs to identify the data points representing the underperforming and outperforming students.

\vspace{3pt}
\begin{wrapfigure}[3]{l}{0.075\linewidth}
    \vspace{-12pt}
    {\includegraphics[height=1.0cm]{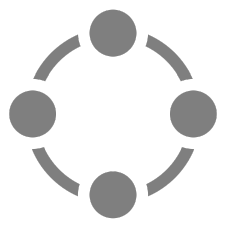}}
\end{wrapfigure}
\noindent\paragraph{N-Annotation Ensembles}
The ensembles used in the collected charts are divided into categories based on the number of individual annotations used in the ensembles, which we identify as 2-annotation, 3-annotation, and 4-annotation ensembles and generally refer to as N-annotations. 
\textit{2-Annotation Ensembles:} Overall, in our study, 2-ensembles were very common, with connector+text, enclosure+text, and glyph+color being the most frequent. For example, in~\autoref{fig:teaser:b__old_d}, the text is used with an arrow (i.e., connector) to present the details of a particular data point in the scatterplot. 
\textit{3-Annotation Ensembles:} 
In our study, the most commonly utilized 3-annotation ensembles were color+enclosure+text and enclosure+connector+text. As illustrated in~\autoref{fig:cs3-b}, color+enclosure+text ensembles were used to identify critical regions of the line chart, where enclosures and color were used to identify the regions, and text was used to present additional context to help the identification task. The enclosure+connector+text ensemble was used when enclosures identified individual or group of data of interest; text presented additional pertinent details for the data point(s), and connectors identified the connection between the data and the related text. An important observation was that when combining four or more distinct annotation types, the annotations often signify redundant or unnecessary encodings. This observation suggests that a more straightforward 2-annotation ensemble might suffice instead of more complex 3- or 4-annotation ensembles.

We further classified N-annotation ensembles into two categories based on the dependency relationship between the individual annotations used in the ensembles.

\vspace{3pt}
\begin{wrapfigure}[3]{l}{0.075\linewidth}
    \vspace{-12pt}
    {\includegraphics[height=1.0cm]{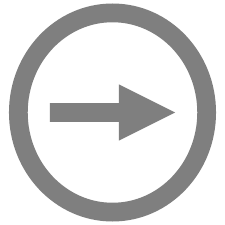}}
\end{wrapfigure}
\noindent\paragraph{One-way} One-way annotation occurs when one annotation can convey its information without relying on another. The secondary annotation merely complements the primary one to enhance the conveyed message. For instance, in \autoref{fig:illustrative-scatterplot:d}, the red and green colors for the cross and circular glyphs do not add new information; they simply emphasize these glyphs.

\vspace{3pt}
\begin{wrapfigure}[3]{l}{0.075\linewidth}
    \vspace{-12pt}
    {\includegraphics[height=1.0cm]{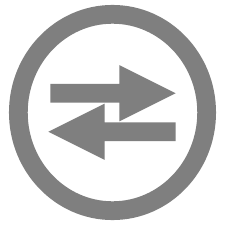}}
\end{wrapfigure}
\noindent\paragraph{Two-way} Two-way annotation ensembles arise when both annotations are interdependent and cannot effectively convey their message in isolation. In \autoref{fig:illustrative-scatterplot:b} and \autoref{fig:illustrative-scatterplot:c}, both the connector and text annotations rely on each other for their intended meanings to be clear. Without either one, the message is unclear.

\subsection{\taskWhat\ --- Classifying Data Sources for Annotations}
\label{sec.what}

Once the specific types or combinations of annotations for an analytical purpose are selected, identifying their data sources becomes essential for their integration into a chart. Aligned with Brehmer and Munzner's framework~\cite{brehmer2013multi}, we define the input to the \taskWhat component as the necessary data for generating annotations, with the annotations themselves being the output. This approach aids in choosing the appropriate data sources for annotation generation. Our analysis of annotated charts indicates that annotations fall into three categories based on their relationship with the data sources that generate them: internal, derived, and external.

\vspace{3pt}
\begin{wrapfigure}[3]{l}{0.075\linewidth}
    \vspace{-12pt}
    {\includegraphics[height=1.0cm]{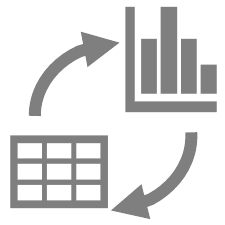}}
\end{wrapfigure}
\noindent\paragraph{Internal Annotations}
Internal annotations refer to annotations that are taken directly from the dataset without requiring any calculations or input from sources outside the dataset. These annotations represent the exact values or text from the dataset. An instance of internal annotations can be observed in \autoref{fig:illustrative-scatterplot:g}, where the text labels associated with each data point represent exact values from the dataset, hence being generated directly from it. Similarly, the text labels in \autoref{fig:teaser:c} for the different point in time on the plotted line and in \autoref{fig:teaser:f} for the pies in the pie chart represent data values taken directly from the dataset, making them internal annotations. 

Internal annotations are prominent, appearing in 973 charts ($\sim$52\%) in our dataset, highlighting their critical role in conveying dataset information to viewers. Text annotations are the most frequent type in this category throughout the dataset, suggesting their popularity in communicating information directly from the dataset to viewers.

\vspace{3pt}
\begin{wrapfigure}[3]{l}{0.075\linewidth}
    \vspace{-12pt}
    {\includegraphics[height=1.0cm]{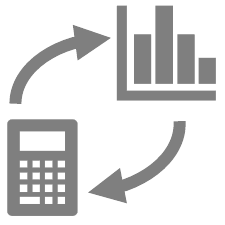}}
\end{wrapfigure}
\noindent\paragraph{Derived Annotations}
Derived annotations are calculated from the data represented by the chart, but do not directly represent values from the dataset. Statistical values and algorithms, such as clustering, are example sources for these types of annotations. 
To clarify, an instance of derived annotations involves calculating and representing a regression, an average, confidence intervals, etc.\ from the chart's dataset. Similarly, applying a clustering algorithm to the dataset and visually representing the clusters through color or enclosure would fall under the domain of derived annotations. In \autoref{fig:illustrative-scatterplot:e}, the indicator (i.e., trend line) depicting the correlation between the two variables is classified as a derived annotation since it was produced by applying an algorithm to the dataset that the chart conveys rather than directly representing any values from the dataset. 

Our analysis shows that derived annotations are comparatively uncommon, appearing in 330 charts ($\sim$17\%), with indicators showing correlations between variables (see~~\autoref{fig:illustrative-scatterplot:e}) being the most frequent within this category. Additionally, a variety of derived annotations were employed to convey statistical information, such as indicators (e.g., vertical lines) in histograms to represent the mean of the distribution. Instances of derived annotations incorporating text, glyphs, and color were also noted across multiple chart types.

\vspace{3pt}
\begin{wrapfigure}[3]{l}{0.075\linewidth}
    \vspace{-12pt}
    {\includegraphics[height=1.0cm]{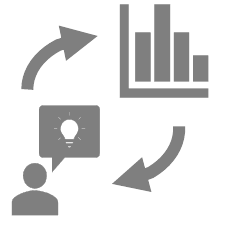}}
\end{wrapfigure}
\noindent\paragraph{External Annotations}
External annotations are those that do not come from the data that created the chart, but instead, they are created using an external data source (e.g., another data file, the internet, information coming from the visualization's author, etc.). For instance, in~\autoref{fig:illustrative-scatterplot}, both the red line and the associated text annotation (see~\autoref{fig:illustrative-scatterplot:f}) combinedly represent the national average score of the given test along the y-axis. Text annotations, such as those in \autoref{fig:teaser:c} and \autoref{fig:teaser:e}, are quite often externally generated. These are considered external annotations as they originate from sources external to the primary data and are added based on knowledge other than that used to generate the scatterplot.

External annotations are most frequently observed in all annotation types, appearing in 994 charts ($\sim$53\%). This widespread application could be due to their flexibility, enabling users to add annotations crucial for articulating the visualization's message or meaning, even when these annotations are not directly related to the data. For instance, users can add text annotations to provide context or explain a data trend, as illustrated in~\autoref{fig:illustrative-scatterplot:c}. \autoref{fig:class-freq-for-annotations} shows the frequency distribution of three categories of annotations for different annotation types.

\begin{table}[!b]
    \centering
    
    \caption{Distribution of various annotation categories based on their data sources across different annotation types, quantified from a total of 3,610 instances. Box colors indicate frequency: \fcolorbox{black}{dsLow}{\tiny 1-5\%} \fcolorbox{black}{dsMed}{\tiny 5-10\%}  \fcolorbox{black}{dsHigh}{\tiny \textcolor{white}{10+\%}}. The last row shows the count of unique occurrences of different annotation categories based on data sources across different annotation types. 
    }
    \label{fig:class-freq-for-annotations}

    \includegraphics[width=0.95\linewidth]{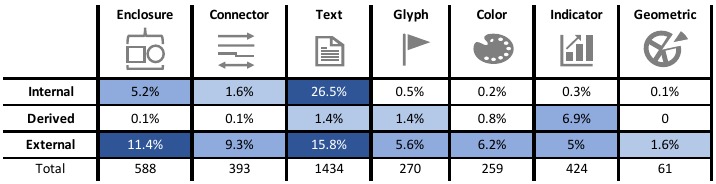}
\end{table}

\section{Examining Design Space Utility Thru Case Studies}
\label{sec.utility}
We conducted three case studies to assess the utility of our proposed design space for annotating visualizations. The case studies aimed to provide examples illustrating how applying the design space improves the understandability and effectiveness of visualizations, demonstrating how it assists in choosing annotations for clear communication across varied visual contexts. In each, we chose a professionally designed annotated chart, removed all annotations to create a baseline, and then reintroduced the annotations, targeting specific communication tasks based on the associated article. This approach to applying annotations was guided by seeking answers to three fundamental questions of our proposed design space in~\autoref{sec.design_space}: \taskWhy aided the annotation process by pinpointing the tasks for which annotations were employed (i.e., \texttt{present}, \texttt{identify}, \texttt{compare}, and \texttt{summarize}); \taskHow facilitated understanding of the methods for implementing annotations (i.e., annotation types and ensembles); and \taskWhat clarified the data informing the creation of annotations (i.e., internal, derived, and external), providing a structured framework for annotating charts.

\subsection{Case Study 1}

\begin{figure*}[!t]
    \centering

    \begin{minipage}[b]{0.92\linewidth}
        \includegraphics[width=0.97\linewidth]{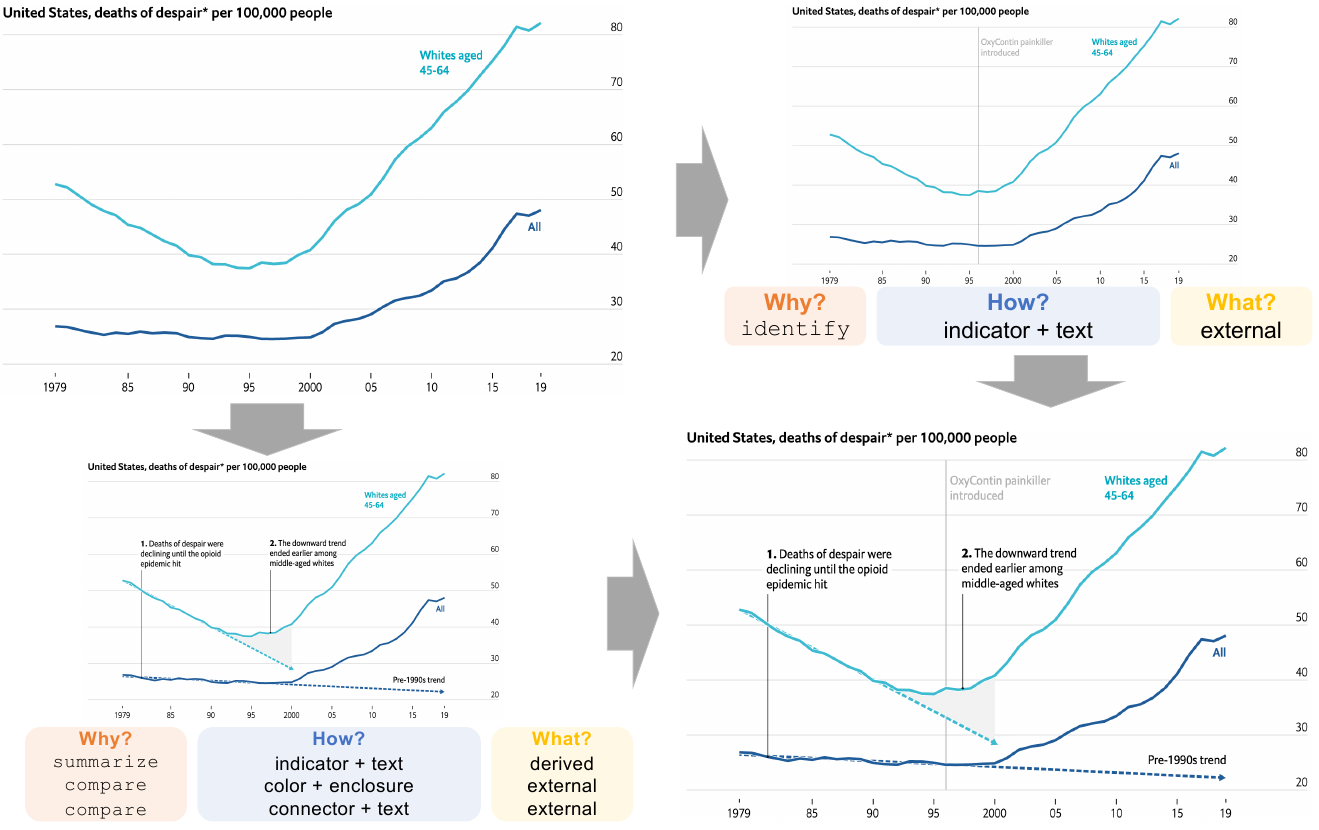}
    \end{minipage}
    \begin{minipage}[b]{1pt}
        \hspace{-508pt}
        \subfloat[\label{fig:cs1-a}]{\hspace{20pt}}
        \hspace{450pt}
        \subfloat[\label{fig:cs1-b}]{\hspace{20pt}}
        \vspace{130pt}
    
        \hspace{-478pt}
        \subfloat[\label{fig:cs1-c}]{\hspace{20pt}}
        \hspace{435pt}
        \subfloat[\label{fig:cs1-d}]{\hspace{20pt}}
        \vspace{130pt}

    \end{minipage}

    \vspace{-5pt}
    \caption{A line chart from \textit{The Economist}~\cite{economist2023religion} depicting factors influencing deaths of despair from 1979 to 2019 on the horizontal axis and the number of deaths on the vertical. (a)~shows the base chart; (b)~uses an indicator+text ensemble to help \texttt{identify} OxyContin's introduction to \texttt{present} its impact on the deaths of despair; (c)~employs indicator+text ensembles to aid in \texttt{summarizing} data trends, and color+enclosure and connector+text ensembles to \texttt{compare} data trends; (d)~presents the fully annotated visualization.}
    \label{fig:case-study-1}
\end{figure*} 
\vspace{-5pt}

In our first case study, we examined a line chart (see~\autoref{fig:case-study-1}), accompanied by an article~\cite{economist2023religion} on factors affecting deaths of despair (i.e., deaths due to drug overdoses, alcohol-related illnesses, and suicide) in the US.

\paragraph{Identifying an External Event (\autoref{fig:cs1-b})} \taskWhy---The first goal was to help \texttt{identify} the introduction of the OxyContin painkiller on the timeline to \texttt{present} the effect of its abuse on the deaths of despair. \taskHow---Our design space indicates that any annotation works for identifying data points in almost all chart types. From these, an indicator (i.e., a vertical line) was selected to help the viewers \texttt{identify} the point in time when the drug was introduced. Understanding that the line might not fully convey the intended message, the indicator was supplemented with text annotations to \texttt{present} additional context about the drug's introduction, aligning with the design space's recommendations for adding additional context. This combination resulted in an indicator+text ensemble, where text provides essential context for interpreting the indicator's significance. \taskWhat---External data for both the indicator and text annotations from the associated article were utilized to represent the introduction of OxyContin, an event outside the line chart's dataset, marking them as external annotations.

\paragraph{Summarizing and Comparing Trends (\autoref{fig:cs1-c})} \taskWhy---Another goal was to help \texttt{summarize} the decreasing trend in deaths of despair preceding the rise of the opioid epidemic in the 1990s and to help \texttt{compare} this decline between whites aged 45-64 and the other demographic group. \taskHow---Indicators are frequently used to summarize data trends across various chart types from the design space. Although indicators (i.e., trend lines) highlight the trend, adding contextual descriptions enhances audience understanding in this context. Consequently, text descriptions were incorporated to reinforce the trend \texttt{summarization} task, creating indicator+text ensembles. Then, for the comparison task, the design space offered several options: enclosures, connectors, text, glyphs, and colors. A color+enclosure ensemble was first chosen to support \texttt{identifying} the area of difference between the trends. Recognizing that color+enclosure highlighting might ambiguously convey the significance of the highlighted area, a connector+text ensemble was introduced to support the \texttt{comparison} task. This ensemble utilizes a connector (i.e., line) to link the area of interest (i.e., the gray highlighted area) with a text explanation, clarifying the premature halt in the declining trend. \taskWhat---The indicators were computed directly from the chart's dataset, identifying them as derived annotations. The text annotations supporting these indicators were based on information not contained within the chart's dataset, thus identifying them as external annotations. Similarly, annotations in color+enclosure and connector+text ensembles, originating from sources outside the dataset, were also identified as external annotations.

\subsection{Case Study 2}

\begin{figure*}[!t]
    \centering

    \begin{minipage}[b]{0.95\linewidth}
        \includegraphics[width=0.97\linewidth]{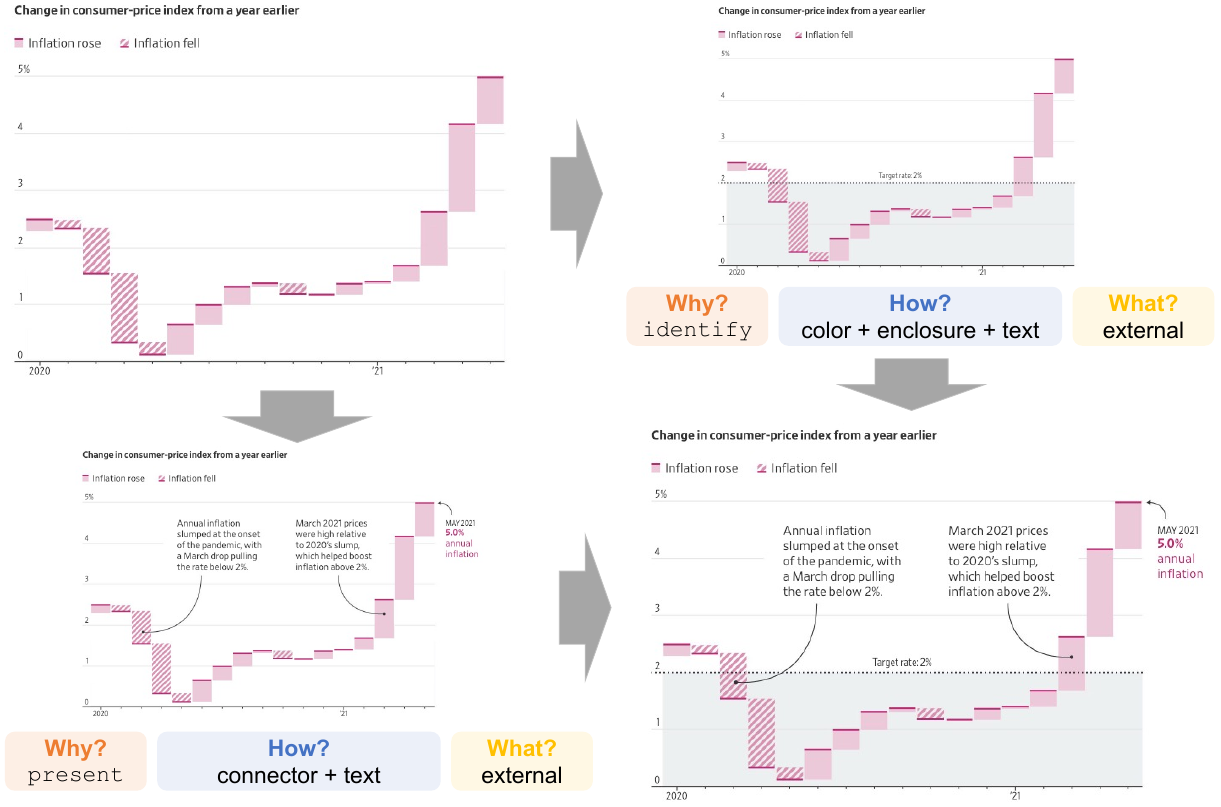}
    \end{minipage}
    \begin{minipage}[b]{1pt}
        \hspace{-508pt}
        \subfloat[\label{fig:cs2-a}]{\hspace{20pt}}
        \hspace{435pt}
        \subfloat[\label{fig:cs2-b}]{\hspace{20pt}}
        \vspace{120pt}
    
        \hspace{-485pt}
        \subfloat[\label{fig:cs2-c}]{\hspace{20pt}}
        \hspace{435pt}
        \subfloat[\label{fig:cs2-d}]{\hspace{20pt}}
        \vspace{100pt}

    \end{minipage}    
    
    \caption{ The bar chart from \textit{The Wall Street Journal}~\cite{DapenaSantilli2021} visualizes inflation fluctuations during the COVID-19 pandemic, with time on the horizontal axis and inflation rate on the vertical. (a)~shows the base chart; (b)~employs a color+enclosure+text ensemble to help \texttt{identify} important chart sections; (c)~uses connector+text ensembles to \texttt{present} article-specific details;  (d)~presents the complete annotated visualization.}
    \label{fig:case-study-2}
\end{figure*}

In our second case study, we investigated a waterfall chart (see~\autoref{fig:case-study-2}), accompanied by an article~\cite{DapenaSantilli2021} that discusses the economic effects of the COVID-19 pandemic on inflation.

\taskWhy---The goal was to help \texttt{identify} the specified threshold (i.e., 2\%) discussed in the associated article and to \texttt{present} the temporary fluctuations in inflation rates below or above the specified threshold due to the pandemic. \taskHow---To assist in \texttt{identifying} the threshold, a color+enclosure+text ensemble was chosen, where the enclosure contained the area below the threshold, the color was used as a highlight to grab viewers' attention, and the text was used to specify further the threshold value (see~\autoref{fig:cs2-b}). Then, to \texttt{present} the fluctuation of the inflation rate, external detail was added using connector+text ensembles according to the design space's recommendation, where the text descriptions present contextual details about the inflation rate fluctuations from the article and connectors help the presentation task by identifying the connection between the text description and the point of interests on the chart (see~\autoref{fig:cs2-c}). \taskWhat---The data for the annotations in both the ensemble types (i.e., color+enclosure+text and connector+text) in~\autoref{fig:cs2-d} came from an external source, distinct from the dataset underlying the baseline chart; therefore, all the annotations are external.

\subsection{Case Study 3}
\label{sec.utility:cs3}

In our third case study, we examined a line chart (see~\autoref{fig:case-study-3}) linked to an article~\cite{Tan2022} regarding the Omicron variant of COVID-19. 

\taskWhy---The goal was to help \texttt{identify} the different peaks of COVID-19 waves and to \texttt{present} additional context from the article about the numbers (i.e., cases, hospitalizations, and deaths) related to the peaks. \taskHow---To aid in \texttt{identifying} the peaks, color+enclosure+text 3-annotation ensembles were used, similar to the second case study, for similar reasoning (see~\autoref{fig:cs3-b}). Then, connector+text ensembles were chosen, according to the recommendation of the design space, to \texttt{present} additional context about the peaks where the text descriptions described additional information regarding the peaks, and the connectors (i.e., arrows) identified the connection between the text descriptions and the points of interest on the chart (see~\autoref{fig:cs3-c}). \taskWhat---Text descriptions, enclosures, connectors, and color highlights that elucidated key aspects of the COVID-19 peaks with information from the associated article, and not from the original dataset, are classified as external annotations. Conversely, certain numerical values within the text descriptions in~\autoref{fig:cs3-c}, taken directly from the dataset, are identified as internal annotations.

\section{Discussion}
\label{sec.discussion}

\subsection{Using the Design Space}
The case studies described in~\autoref{sec.utility} illustrate the application of our proposed design space in annotating charts. Further uses of the design space are the following:

\paragraph{A Structured Way to Think About Constructing and Critiquing Annotations} By combining common practices for data, task, and encoding, our design space enables individual authoring annotations to decide, in an organized way, what annotations to implement. Further, the design space provides a structured method for critiquing others' use of annotations by identifying and categorizing the key elements used in their annotation.

\paragraph{Opening Up the Space of Annotations} Despite annotations being a tool used nearly every day, the space of possible annotations encodings available was not initially obvious to us. The design space provides a comprehensive list of annotation types to apply to visualizations. Importantly, whether intentional or not, the annotations we observed participants using within this design space seemed closely linked to the idea of encoding semantics~\cite{ware2019information}.

\paragraph{Considering Common Practices} Our design space in \autoref{fig:design-space} highlights the annotation types commonly associated with specific visualization types. Practitioners can gain insights into prevalent practices by observing frequently and infrequently used combinations. This understanding encourages thoughtful consideration of commonly used annotation strategies to enhance communication. Additionally, our findings can potentially aid in developing tools to facilitate appropriate annotations for different chart types.

\subsection{Ecological Validity and Limitation}

Our study's ecological validity is supported by the analysis of 14 types of static charts sourced from a diverse range of real-world contexts. This approach lends some weight to the generalizability of our proposed design space for chart annotations. We acknowledge that further empirical studies are needed for complete validation, but the diversity in our dataset provides a reasonable basis for believing that our design space could be applicable in a variety of scenarios.

However, several limitations exist. First, the absence of information about the annotators’ expertise and the tools they used presents a gap in our data understanding. Furthermore, the lack of direct communication with designers made it challenging to fully understand the narrative goals behind annotation use. Research also shows that  visualization goals often misalign with viewer comprehension, influenced by factors such as visualization features, individual backgrounds, graph complexity, etc.~\cite{Quadri_2024}. Therefore, even if designers have clear communicative goals (e.g., to identify, summarize, or compare), the viewers may not necessarily interpret the visualizations as intended. Second, our taxonomy and design space for chart annotations are based on qualitative analysis, and we have yet to evaluate the effectiveness of different annotations across various chart types, preventing definitive claims about their efficiency. Additionally, our study focuses on static visualizations. Dynamic, interactive, and less frequent chart types were not explored and may expand the design space further.

\subsection{Future Work}
\vspace{-3pt}
To build upon our study, future research should address several identified limitations and explore new avenues for insight. A critical next
step is the empirical evaluation of different annotation types across
various chart types, adding a quantitative dimension to our qualitative
taxonomy and design space, enabling precise conclusions on annotation
effectiveness in diverse scenarios. Moreover, with the rise of interactive
visualizations as essential tools for real-time data analysis and nuanced
interpretation, investigating how annotations perform in such dynamic
environments is critical to determine if our current design principles
hold or if new categories emerge specific to interactivity. Additionally,
the insights of visualization professionals who utilize annotations in
various contexts are invaluable; exploring their practices and viewpoints
will illuminate real-world challenges and usage scenarios of annotations in visualizations, enriching our understanding and application of
annotation strategies. Finally, future studies should examine the alignment between designers' communicative goals with annotations and audience interpretation, as understanding this relationship can bridge the gap between annotation creation and viewer comprehension.

\section{Conclusions}
\vspace{-3pt}
Annotations are crucial for visual data exploration, serving as critical aids that facilitate hypothesis generation, information communication, and sensemaking while analyzing charts. Understanding the usage patterns of annotations in visualizations can provide design guidelines for creating compelling and expressive visualizations. We proposed a design space for annotations applicable to diverse chart types, offering practical insights on different annotation types, their common usages, task-annotation pairings, annotation ensembles, and data sources for annotations. Through our design space, we encourage the development of effective and efficient visualizations that can support better user engagement and comprehension.

\acknowledgments{
We used the generative AI tool ChatGPT 3.5 by OpenAI to edit, paraphrase, and restructure the text, which was later reviewed and revised. Figures from outside sources have been cited and used in accordance with the fair use doctrine. This work was partially supported by NSF IIS-2316496, DUE-2216227, and by CNS-2127309
to the CRA for the CIFellows Program.
}

\setstretch{0.975}
\bibliographystyle{abbrv-doi-hyperref}

\end{document}